\documentclass[12pt,preprint]{aastex}

\shorttitle{Hinode Giant Chromospheric Jet}
\shortauthors{Nishizuka et al.}

\begin{document}

\title{Giant Chromospheric Anemone Jet observed with Hinode and Comparison with
Magnetohydrodynamic Simulations: \ 
Evidence of Propagating Alfv$\acute{e}$n Waves and Magnetic Reconnection}


\author{N. Nishizuka\altaffilmark{1}, M. Shimizu\altaffilmark{2}, 
T. Nakamura\altaffilmark{1}, K. Otsuji\altaffilmark{1}, 
T. J. Okamoto\altaffilmark{3}, Y. Katsukawa\altaffilmark{3} 
and K. Shibata\altaffilmark{1}}


\altaffiltext{1}{Kwasan and Hida observatories, Kyoto University, Yamashina, Kyoto, 607-8471, Japan; nisizuka@kwasan.kyoto-u.ac.jp}
\altaffiltext{2}{Department of Mechanical Engineering and Science Graduate School of Engineering, Kyoto University, Sakyo-ku, Kyoto 606-8501, Japan}
\altaffiltext{3}{National Astronomical Observatory of Japan, Mitaka, Tokyo, 181-8588, Japan}

\begin{abstract}
Hinode discovered a beautiful giant jet with both cool and hot components at the solar limb on 2007 February 9. 
Simultaneous observations by the Hinode SOT, XRT, and TRACE 195 $\AA$ satellites revealed that hot 
($\sim 5\times10^6$ K) and cool ($\sim10^4$ K) jets were located side by side and that the hot jet preceded the 
associated cool jet ($\sim1-2$ min.). A current-sheet-like structure was seen in optical (Ca $_{II}$ H), 
EUV (195 $\AA$), and soft X-ray emissions, suggesting that magnetic reconnection is occurring in the transition region 
or upper chromosphere. Alfv$\acute{e}$n waves were also observed with Hinode SOT. These propagated along 
the jet at velocities of $\sim 200$ km s$^{-1}$ with amplitudes (transverse velocity) of 
$\sim$5-15 km s$^{-1}$ and a period of $\sim200 s$.
We performed two-dimensional MHD simulation of the jets on the basis of the emerging flux - reconnection model, 
by extending Yokoyama and Shibata's model. We extended the model with a more realistic initial condition 
($\sim10^6$ K corona) and compared our model with multi-wavelength observations. 
The improvement of the coronal temperature and density in the simulation model allowed for the first time 
the reproduction of the structure and evolution of both the cool and hot jets quantitatively, supporting the magnetic 
reconnection model. The generation and the propagation of Alfv$\acute{e}$n waves are also reproduced 
self-consistently in the simulation model.
\\
\end{abstract}




\section{Introduction}
The solar chromosphere has been known to be very dynamic \citep[e.g.,][]{bray74,zir88}. However, recent 
Hinode observations \citep{kos07} revealed that it is even more dynamic than previously thought 
\citep{shi07,kat07,cir07}(see also special issue, Initial Results from 
Hinode, 2007, PASJ, Vol. 59, No. SP3). It has long been observed that H$\alpha$ jets called surges often occur in the 
chromosphere \citep[see e.g.][]{rus67,kur92}. Surges are believed to be produced by magnetic 
reconnection \citep{hey77,yok95,yok96,iso05}, which is an energy conversion mechanism from magnetic 
energy into thermal and kinetic energies of plasma when two anti-parallel magnetic fields encounter 
and reconnect with each other. New chromospheric observations with the calcium $_{II}$ H-line broad 
band filter of the Solar Optical Telescope \citep[SOT;][]{tsu08} on board Hinode revealed that jets are 
ubiquitous in the chromosphere and some of the jets show evidence of magnetic reconnection \citep{shi07,kat07}. 

There have been several multi-wavelength observations of jets (e.g. X-ray and H$\alpha$ \citep{sch95,can96}, 
H$\alpha$ and EUV \citep{chae99,jian07}, and EUV and X-ray \citep{alex99}). These have shown that 
hot jets (X-ray or EUV jet) and cool jets (H$\alpha$ surge or dark EUV jet) are almost co-spatial 
and co-temporal, and that they are dynamically connected to each other. 
Yokoyama and Shibata (1995, 1996) explained such a relation between hot and cool jets qualitatively by 
performing a resistive MHD simulation of an emerging flux model. However, their model was not fully realistic 
because of various computational difficulties; it was not possible to reproduce both 
emerging flux and jets self-consistently in one model, so they assumed an unrealistic initial 
temperature ($\sim2.5\times10^5$ K) and density ($10^{12}$ cm$^{-3}$ corona) for the corona.   
Because of this, the coronal Alfv$\acute{e}$n speed in their model was lower than the actual value, 
so the jet velocity was also lower than the observed velocity of the H$\alpha$ surge/jet.
In this Letter,  we succeed in modeling for the first time both the emerging flux and jet self-consistently with 
realistic coronal temperatures ($\sim10^6$ K) and densities and attempt to compare simulation results 
and observations quantitatively.
\section{The 2007 February 9 Giant Ca Jet Event}
The solar jet studied here occurred on the west limb of the Sun around NOAA active region 
10940 on 2007 February 9 at 13:20 UT, which was associated with a weak soft X-ray brightening observed 
in GOES. The maximum height of the jet was $\sim14,000$ km and its 
width $\sim6,000$ km. Figure 1a-1c shows Ca $_{II}$ H-line broad-band filter snapshot images 
taken with SOT on board Hinode, which provide a high spatial resolution (0.2" or 
150 km on the solar surface) and stable observation of the photosphere and chromosphere. 
The cold plasma jet, which we call the Ca jet, began to be ejected upward with inclination 
angle of 45 degree at 13:18 UT. The jet grew up to a cusp- or inverted-Y shaped structure, 
whose morphology was similar to that of coronal anemone jets \citep{shi94}. 
The maximum upward velocity occurred 10 minutes later at 13:30 UT (also see Fig.2a). 
The jet was accelerated to $\sim100$ km s$^{-1}$ along the field line. The cool plasma subsequently 
fell down with a decelerated free-fall motion.

Yohkoh observations have led to the interpretation of the anemone-shaped structure as a result of magnetic reconnection between 
an emerging magnetic bipole and a preexisting coronal uniform field \citep{shi92,shi94,shimj96}. 
This interpretation has been supported by magnetohydrodynamic (MHD) simulations of emerging flux 
\citep{hey77,yok95,yok96,iso05}. By extending previous simulations the new simulations consider a more realistic 
initial condition, which allow a comparison with observations.
\footnote{
Initial coronal and chromospheric/photospheric temperatures are 100 and unity in unit of $10^{4}$ K, 
respectively, and plasma $\beta$ (= ratio of gas pressure to magnetic pressure) is 4 in the horizontal 
flux sheet just below the photosphere, 0.01-0.06 in the inner corona, 
and varies from 4 to 0.06 in the chromosphere/photosphere. Note that this simulation 
is based on an idealized model, in which there is no difference between chromosphere and photosphere 
in temperature. 
Numerical computation was carried out using the CIP-MOCCT scheme \citep{kud99} with total grid points 
of ($2000\times1000$), whose grid size is 0.2 ($\sim40$ km) and uniform in the whole 
computational box. We assumed anomalous resistivity \citep{ugai85}, whose functional 
form is $\eta=0$ for $v_{d}<v_{c}$ and $\eta\propto(v_{d}/v_{c}-1)^{2}$ for $v_{d}>v_{c}$, where 
$v_{d}\equiv J/\rho$, is the drift velocity, $\rho$ is the mass density, $J$ is the current density and 
$v_{c}$ is the threshold above which anomalous resistivity sets in.
Typical magnetic Reynolds number $V_A L/\eta$ at the location of reconnection is 
$\sim 1000$, where $V_A$ is the Alfv$\acute{e}$n speed just outside the current sheet, $L$ is the length of the 
current sheet, and $\eta$ is the magnetic diffusivity.
The magnetic Reynolds number used in our simulation model is much lower than the actual number for solar corona, 
and the assumed functional form for the anomalous resistivity has not yet been physically established.
However, it should be noted that this type of anomalous resistivity reproduce well the fast reconnection model 
\citep[e.g.,][]{ugai85} and explosive energy release in solar flares \citep[e.g.,][]{yok94}. 
}
Our initial condition of the simulations is basically similar to that of Yokoyama \& Shibata (1995), 
with the only difference being in coronal temperature and density; they were assumed to be $2.5\times10^{5}$ K 
and $10^{12}$ cm$^{-3}$ in the Yokoyama-Shibata model, but $10^{6}$ K and $10^{10}$ cm$^{-3}$ in our model. 
Note that our coronal parameters are much more realistic than theirs. 

As a result of our new two-dimensional MHD simulation, we found that a simulated jet is amazingly 
similar to the observed jet. The comparison among them is shown in Figure 1 (d-f are simulation 
results). These simulations are performed by solving the two-dimensional resistive MHD equations with 
uniform gravitational field and without thermal conduction and radiative cooling effects. 
We note that a jet is magnetically driven, not gas pressure driven, so the 
dynamics of the jet cannot be influenced by thermal conduction and radiative cooling effects even if the temperature 
is not fully realistic. We also note that evaporation effects due to thermal conduction are not included 
in our model, so the evaporation-driven X-ray jets \citep{simj01,miy04} cannot be modeled in the Letter. 
As an initial state, we set hydrostatic plasma in the corona ($10^{6}$K), chromosphere/photosphere 
($10^{4}$K), and convection zone with a horizontal flux sheet just below the photosphere and a uniform oblique 
magnetic field outside the flux sheet in the whole region. This flux sheet is unstable for the magnetic 
buoyancy instability \citep[the Parker instability;][]{shi89,mat93}, so the perturbed flux sheet is 
excited and emerges into the corona. 
While we performed a two-dimensional simulation, the emerging process of magnetic 
flux would actually become much more complex in three dimensions \citep[see e.g.][]{fan03,iso05}.
In two dimensions, once reconnection occurs between the emerging flux and preexisting 
field, the field lines with a polarity opposite to that of the ambient field become connected to the ambient 
polarity regions, forming an inverted-Y shape or anemone jet. Because the time and spatial scales are 
arranged almost in the same way as in Figure 1a-c, the simulation model turns out to be able to explain 
observational facts very well. Reconnection creates multiple islands which confine cool, 
dense, chromospheric plasma in the current sheet, which are ejected at the Alfv$\acute{e}$n 
speed $v_{A}\sim 10^2$ km s$^{-1}$ (B/20G)(n/$10^{12}$ cm$^{-3})^{-1/2}$$\sim$50-150 km s$^{-1}$ 
if we assume coronal field B$\sim$20 G and density n$\sim10^{11-12}$ cm$^{-3}$ in the upper chromosphere, in 
agreement with the observations (shown in Fig.2a). These facts support a magnetic reconnection 
model. Note that, if reconnection occurs below the lower chromosphere, shocks are formed in 
front of a jet due to rapid decrease of plasma density \citep[e.g.,][]{shi82,tar99}, which may eventually 
accelerate jets along magnetic field lines. However, in such a case, only cool jets are formed. In our case, 
it seems that reconnection occurred in the transition region or upper chromosphere, because not only cool 
jets but also hot jets are observed simultaneously (see section 3).

Figure 2b shows the distance-time diagram of the position of the Ca jet (as represented by Ca intensity 
distribution), which revealed the oscillation of the jet with amplitude of 5-15 km s$^{-1}$ and period of 
200 s.  From the oscillation pattern at three different heights, we find that the oscillation propagates 
along the jet at 100-200 km s$^{-1}$. Since the jet is believed to be along the magnetic field, the propagation 
of the oscillation is the evidence of the propagating Alfv$\acute{e}$n waves. The simulation model reproduced 
also the generation and propagation of Alfv$\acute{e}$n waves (see plus symbols in Fig. 2b).  Hinode 
observations revealed the existence of Alfv$\acute{e}$n waves or Alfv$\acute{e}$n-like oscillation 
\citep{oka07, dep07}, but {\it propagating} Alfv$\acute{e}$n waves have not been observed until now. 
Hence our observations are the first observational evidence suggesting the presence of {\it propagating} 
Alfv$\acute{e}$n waves.




\section{Comparison between Hot/Cool components of the Jet and MHD simulation}

Our simulation shows that both hot and cool jets can be accelerated simultaneously by 
magnetic reconnection driven by emerging flux, as in Yokoyama \& Shibata (1995). 
The plasma in the corona is heated to temperatures from a few million K to about 10 
million K. This hot plasma can be observed as microflares and soft X-ray jets. At the same 
time, cool jets are accelerated if the reconnection occurs in the transition region or in 
the upper chromosphere, where cool plasma is situated near the reconnection point. According 
to the model, X-ray or EUV jets are seen as hot jets that are accelerated by the magnetic 
tension force of the reconnected field lines, and H$\alpha$ surges or Ca jets are seen as cool 
jets that are accelerated by the slingshot effect due to the reconnection, which produces 
a whiplike motion. Furthermore, if a large amount of energy is injected into the upper 
chromosphere, the cold plasma above the energy injection point is heated up and evolves into 
evaporation \citep{simj01}, although we cannot reproduce it because of the lack of thermal 
conduction in the energy equation. Such a coexistence of hot and cool jets is indeed confirmed 
by observations with the comparisons between associated X-ray or EUV jets and H$\alpha$ surges 
\citep{sch95,can96,chae99,alex99,ko05,jian07}. 

Figure 3a-o shows the time evolution images of the jet with multiwavelength 
observations taken with the SOT and X-Ray Telescope \citep[XRT;][]{gol07} aboard Hinode, and the Transition 
Region and Coronal Explorer \citep[TRACE;][]{handy99} 195$\AA$ filter. An X-ray image before ejection 
shows a loop structure over the limb, at the footpoint of which a jetlike feature appeared at 13:09 
UT, preceding other wavelengths. X-ray emission gradually increased its intensity and, 
just after 13:16 UT, an X-ray jet started to be ejected, and cusp- or inverted-Y shaped structure 
was formed. An EUV jet was ejected from a looplike bright emission patch at the moment when the patch 
reached its maximum intensity. The cold plasma ejection, such as the Ca jet and a dark EUV jet in 
absorption ($\sim 10^{4}$ K), was delayed by 1-2 minutes relative to the X-ray and EUV brightening. The 
maximum upward velocity of the cool jets occurred about 10 minutes later than the X-ray spike (also see 
Fig.2a). As a result of spatial co-alignment, the EUV jet was identified with the X-ray 
jet, whereas the dark EUV jet appeared to be a counterpart to the Ca jet. 
The X-ray jet and the Ca jet were ejected side by side with each other, which is the same feature as shown 
in the simulation results of Figure 1i. These observational facts indicate the following 
three features: (1) The X-ray jet ($5\times 10^{6}$ K) and the EUV jet ($\sim 10^{6}$ K) 
are likely to have the same physical origin, and the Ca jet ($10^{4}$ K) and the dark EUV jets (unknown 
temperature, $\sim 10^{4}$ K) also have the same origin. (2) The X-ray and the Ca jets are different 
kinds of plasma ejection, implying dynamically connected hot and cool plasma ejections along different 
field lines. (3) The X-ray jet precedes the Ca jet by 1-2 minutes compared with their maximum 
intensity time, and some hot structure was also observed with X-ray emission 10 minutes earlier than 
jet ejection.

We note that the delay of the cool jet may not be due to a cooling effect, because the delay 
time is much shorter than the cooling time. Furthermore our simulation without a cooling term reproduces both hot 
and cool jet very well. According to the simulation, the current sheet is formed between the corona (hot and 
thin plasma) and the emerging flux with chromospheric density and temperature (cool and dense plasma). 
Hence the reconnection is very asymmetric \citep{pet67}. In such a case, the low-density part becomes a hot 
and fast jet, while the high-density part forms a cool and slow jet, because the local Alfv$\acute{e}$n speed 
is high (low) in the low (high) density part. This is why a hot jet preceded a cool jet; i.e., the hot jet 
reaches higher altitude earlier than the cool jet.

\section{Summary and Discussion}

MHD simulation results (Fig.1d-i) reproduce remarkably well the dynamics 
and structure of cool and hot jets and their relative timings. The only structure that the simulations cannot 
explain is the existence of an X-ray bright point (probably an unresolved loop) in Fig.3m-o. 
However, MHD simulations in Figure 1 are two-dimensional, and hence discrepancy between observations 
and simulations may be explained by three-dimensional effects. In fact, the reconnection model \citep{yok95,
yok96,shi92,shi94,shimj96} predicts the formation of not only jets but also loops in a separated place, 
which can explain many X-ray observations showing that bright points (loops) are situated separately from 
jets \citep{shi92,shi94,shimj96}. In our case, such bright loops may be situated just in front of jets 
because of three-dimensional projection effects.  

On the other hand, detailed comparison between simulations and observations suggests that the 
current sheet structure in Figure 1e may be visible as one of legs in the inverted-Y shaped 
structure. It is interesting to see that EUV and X-ray loops seem to be situated along the 
same leg, i.e., possibly corresponding to the current sheet.

We note here that an X-ray brightening at the footpoint cannot be seen in EUV emission 
but in absorption. This may be interpreted as either the three-dimensional effect that the 
EUV emission from the X-ray source is covered by cool plasma ejection or the temperature effect 
such that the temperature in the X-ray source is too high to emit enough EUV emission. 
(Note that usually X-ray loops cannot be seen in EUV images.)

It is also noted that both observations and simulations show the generation of Alfv$\acute{e}$n waves when 
magnetic reconnection occurs: the jet undergoes apparent motion perpendicular to the jet direction at 
amplitudes of 5-15 km s$^{-1}$ (see Fig. 2b). This velocity is also comparable to that 
observed for polar X-ray jets \citep{cir07}. These Alfv$\acute{e}$n waves 
are generated by reconnection \citep{yok98,tak01}, and 
may contribute to the heating and acceleration of solar wind when the magnetic field is open 
\citep{par88,suz05}.



\acknowledgments

We thank T. Yokoyama and M. Shimojo for useful comments. Hinode is a Japanese mission developed and launched by ISAS/JAXA, 
with NAOJ as domestic partner and NASA and STFC (UK) as international partners. It is operated by these 
agencies in co-operation with ESA and NSC (Norway). TRACE is a NASA Small Explorer mission. This work is 
supported in part by the Grant-in-Aid for the global COE program "The Next Generation of Physics, Spun from University 
and Emergence" from the Ministry of Education, Culture, Sports, Science, and Technology (MEXT) of Japan, and also in 
part by the Grand-in-Aid for Creative Scientific Research ``The Basic Study of Space Weather 
Prediction'' (Head Investigator: K.Shibata) from the Ministry of Education, Culture, Sports, Science, and 
Technology of Japan. The numerical computation was performed on Fujitsu VPP5000 at NAOJ.\\



\clearpage




\begin{figure}
\plotone{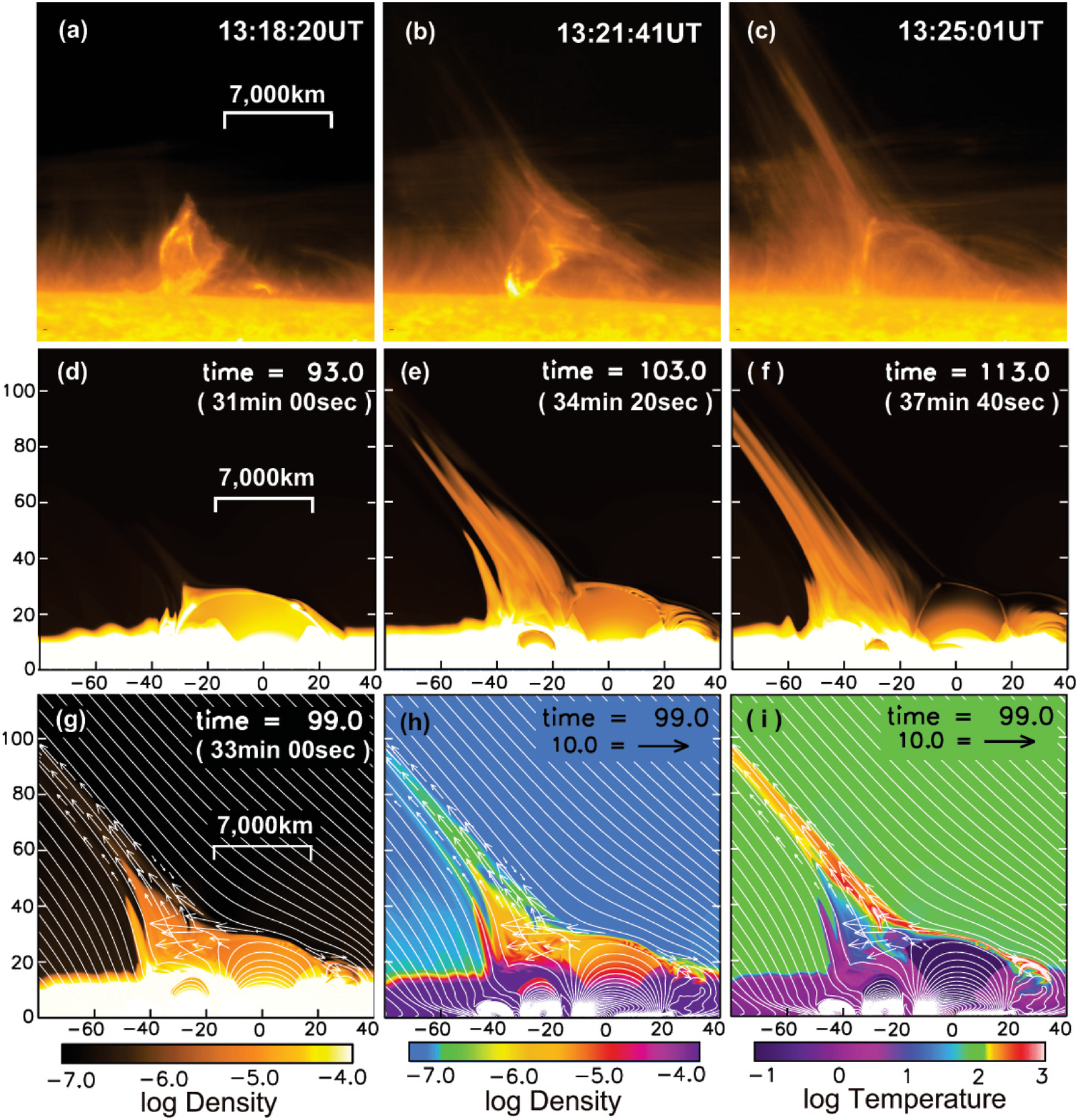}
\caption{Comparison between the Ca jet and simulated jets based on the reconnection model. (a)-(c) Ca II H 
broadband filter images of the Ca jet on 9 February 2007, taken with Hinode/SOT. (d)-(f) Two-dimensional 
distributions of density (log$_{10}$ $\rho$; color map) in units of 10$^{-7}$ g cm$^{-3}$ 
in the simulated jets, whose times correspond to those of the observed jets in the top panels. 
The length is in units of 200 km and the times are in units of 20 s.
(g-i) Two- dimensional distribution in density (log$_{10}$ $\rho$; color maps) in units of 
$10^{-7}$ g cm$^{-3}$ and temperature (log$_{10}$ T; color map) in units of $10^4$ K of magnetic fields 
(B; lines) and velocity vectors (v; arrows) in the simulated jets. \label{fig1}}
\end{figure}


\begin{figure}
\epsscale{.68}
\plotone{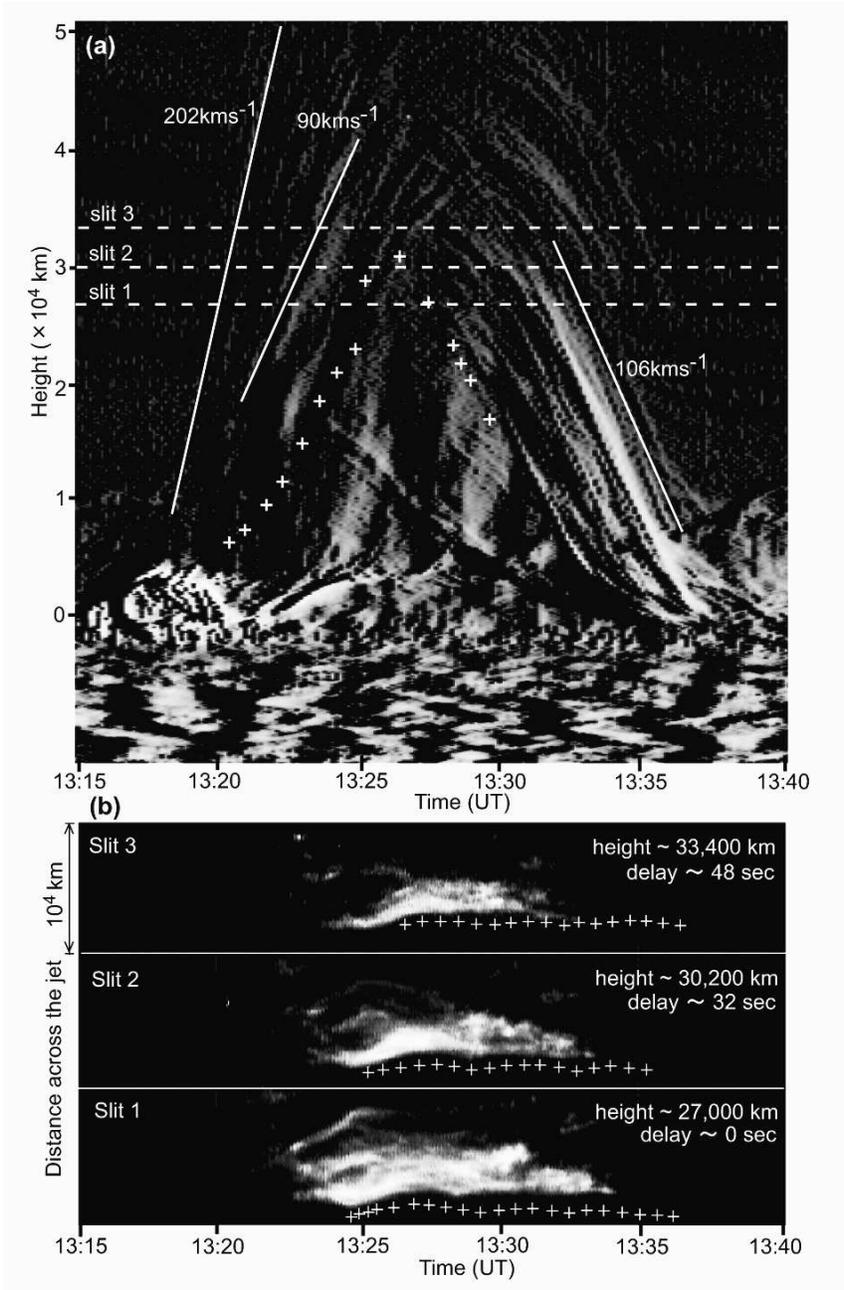}
\caption{
(a) Height-time diagram of intensity (running differences) along the Ca jet taken with Ca $_{II}$ H 
broadband filter of Hinode SOT. The numbers above the curves denote velocity (in km s$^{-1}$). The plus symbols 
show the height of the simulated jet shown in Fig. 1. The simulation reproduced well the observed motion of 
the jet. (b) Distance-time diagram of Ca intensity across the jet, revealing the oscillation 
of the jet at amplitude of 5-15 km s$^{-1}$ and period of 200 s, at three locations (heights). It is found 
that the oscillation propagates from lower altitude at slit 1 position to higher altitude at slit 2 and 3 
positions. Since the jet is considered to be parallel to magnetic field lines, this is the evidence of the 
Alfv$\acute{e}$n wave propagating at the velocity 100-200 km s$^{-1}$.  The simulated jet also shows the 
generation and propagation of Alfv$\acute{e}$n waves along the cool jet, which is indicated by the plus symbols
(the position of the jet).\label{fig2}}
\end{figure}

\begin{figure}
\epsscale{.65}
\plotone{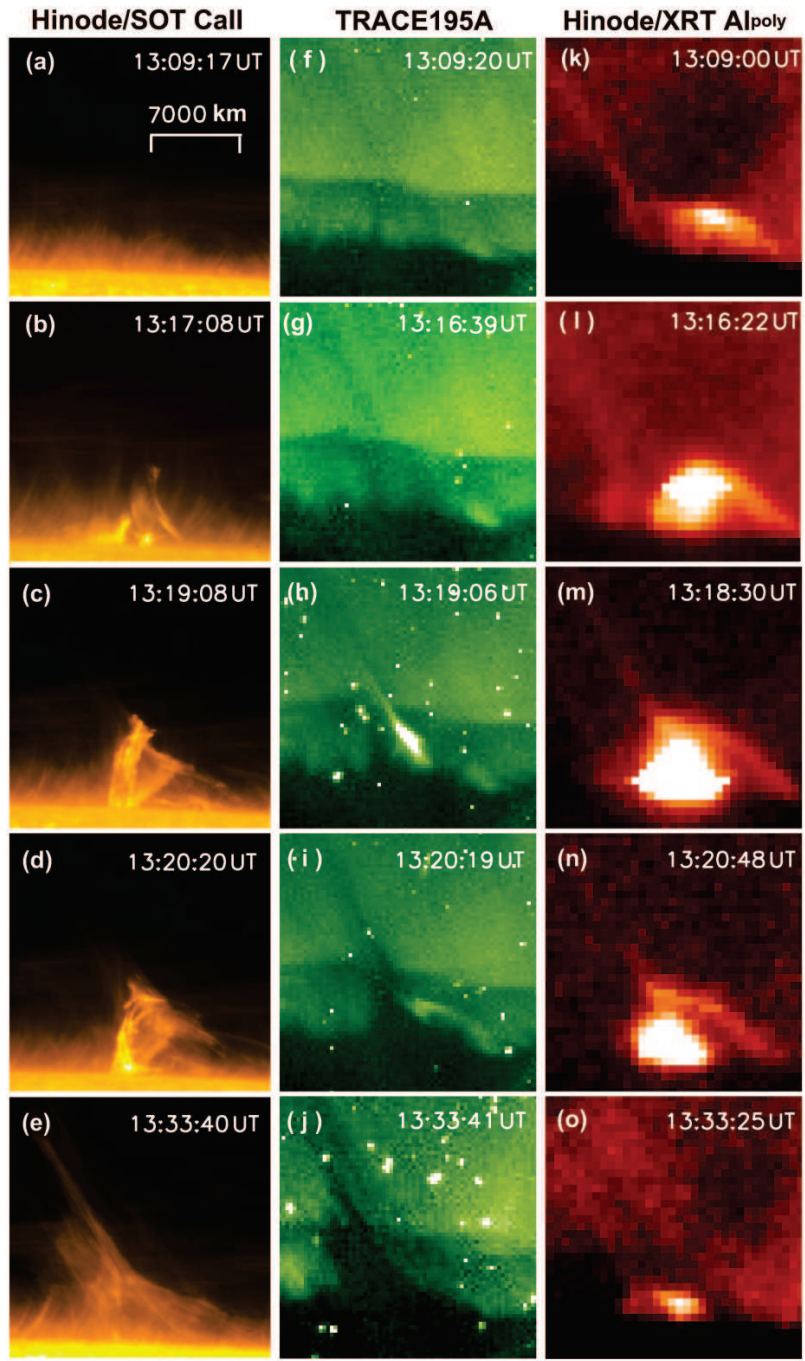}
\caption{Comparison of multi-wavelength observations. 
(a-e, f-j, k-o) Time evolution images of the Ca jet taken with Hinode/SOT Ca II H-line broad 
band filter, the EUV jet with TRACE/195$\AA$ filter and the X-ray jet with Hinode/XRT Al poly filter. 
\label{fig3}}
\end{figure}




\end{document}